\documentclass[fleqn,10pt,preprint,1p]{elsarticle}
\usepackage[intlimits,reqno]{amsmath}
\usepackage{cite}
\biboptions{sort&compress}
\def\oo{\infty}
\def\Z#1{\zeta(#1)}
\def\A#1{a_{#1}}
\def\B#1{b_{#1}}
\def\D#1{d_{#1}}
\def\Re{\mathrm{Re}}
\def\Im{\mathrm{Im}}

\def\app{{\left(\frac{\alpha}{\pi}\right)}}
\def\SYS{{\tt SYS}}
\def\Li{{\mathrm{Li}}}
\def\cl#1#2{{\mathrm{Cl}_{#1}}\left({#2}\right)}
\def\Cl#1{{\mathrm{Cl}_{#1}\left(\frac{\pi}{3}\right)}}
\def\Cldd{{\mathrm{Cl^2_2}\left(\frac{\pi}{3}\right)}}
\def\cat#1{{\mathrm{Cl}_{#1}\left(\frac{\pi}{2}\right)}}
\def\catdd{{\mathrm{Cl}^2_{2}\left(\frac{\pi}{2}\right)}}
\def\ZTD{\zeta^2(3)}
\def\rha#1{\Re H_{#1}\left(e^{i\frac{\pi}{3}}\right)}
\def\rhb#1{\Re H_{#1}\left(e^{i\frac{2\pi}{3}}\right)}
\def\iha#1{\Im H_{#1}\left(e^{i\frac{\pi}{3}}\right)}
\def\ihb#1{\Im H_{#1}\left(e^{i\frac{2\pi}{3}}\right)}
\def\rhe#1{\Re H_{#1}\left(e^{i\frac{\pi}{2}}\right)}
\def\ihe#1{\Im H_{#1}\left(e^{i\frac{\pi}{2}}\right)}
\def\rs{\sqrt{s}}
\def\btell{B_3}
\def\ctell{C_3}
\def\WF{{\tilde{F}}} 
\def\intaxoneones#1{f_1(#1)}
\def\intaxonetwos#1{f_2(#1)}

\def\aql#1{A_{#1}}
\def\cql#1{a_e^{(#1)}}

\def\e{\epsilon}

\def\Fcub#1{{}_2F_1\left(\begin{smallmatrix}
{{ \frac{1}{3}\;\frac{2}{3} }}\\
{{1 }}\end{smallmatrix}; #1\right)}
\def\Ecub#1{{}_2F_1\left(\begin{smallmatrix}
{{ \frac{1}{3}\;-\frac{1}{3} }}\\
{{1 }}\end{smallmatrix}; #1\right)}
\def\TITLE{High-precision calculation of the 4-loop QED contribution
to the slope of the Dirac form factor}

\def\CAPTIONFIGA{The 4-loop self-mass diagrams.}
\def\CAPTIONFIGB{The 25 gauge-invariant sets. We show one single
vertex diagram for each set.}
\def\CAPTIONFIGC{Master integrals known only numerically.
$(f,f',f'')$ and $(g,g',g'')$ have numerators equal to $(1,p.k,(p.k)^2)$, respectively.} 
\def\CAPTIONTABLEA{First 1100 digits of $\aql{4}$.}
\def\CAPTIONTABLEB{Contribution to $\aql{4}$ of the 25 gauge-invariant sets of Fig.\ref{figuragau}.}
\def\CAPTIONTABLEC{Numerical values of the constants appearing in Eq.\ref{Tall}.}
\def\eqref#1{Eq.(\ref{#1})}
\def\eqrefb#1#2{Eqs.(\ref{#1})-(\ref{#2})}

\def\comba#1{{t_#1}}
\def\combp#1{{v_#1}}
\def\combel#1{{e_#1}}
\def\Figuraboh 
{ 
\begin{figure} 
\begin{center}
\includegraphics[scale=0.41]{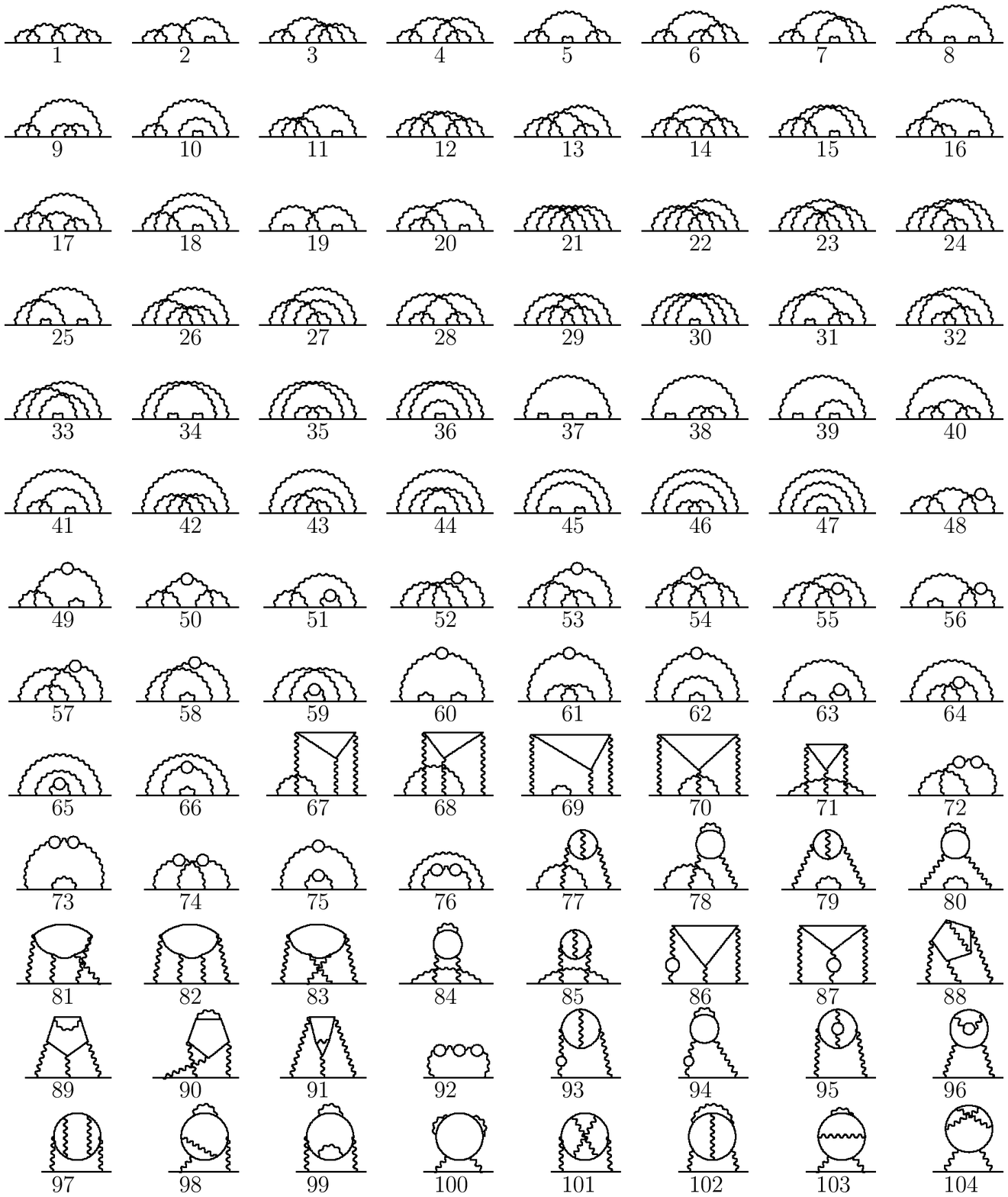}
\caption{\CAPTIONFIGA}
 \label{self104}
\end{center}
 \end{figure} 
 }            
\def\Figuragau 
{ 
\begin{figure} 
\begin{center}
\includegraphics[scale=0.625]{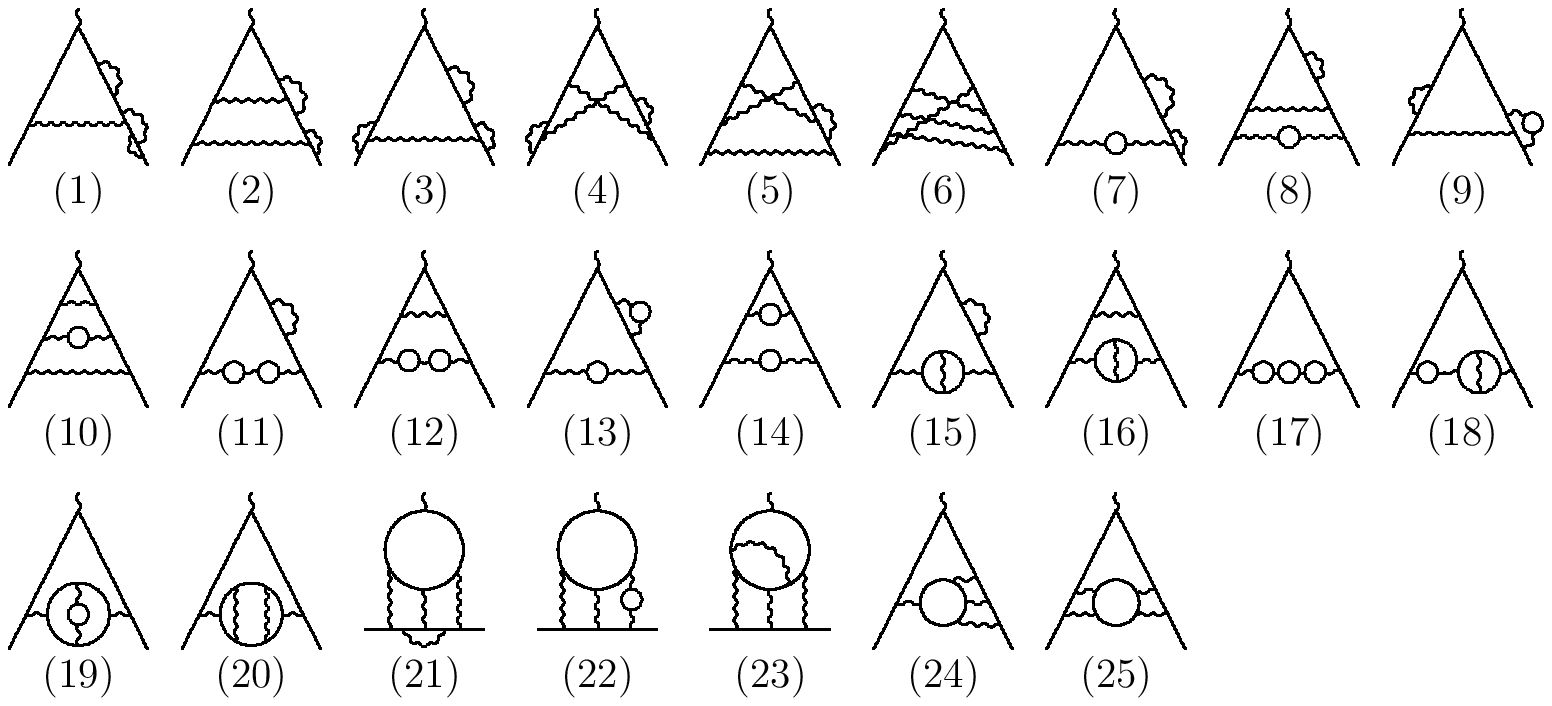}
\caption{\CAPTIONFIGB}
\label{figuragau}
\end{center}
 \end{figure} 
 }            
\def\Figuraunkmasd
{ 
\begin{figure} 
\begin{center}
\includegraphics[scale=0.875]{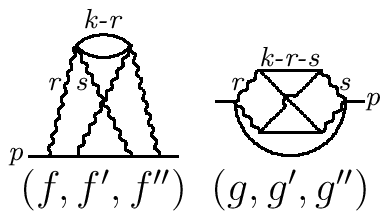}
\caption{\CAPTIONFIGC}
 \label{unkmasd}
\end{center}
 \end{figure} 
 }            
\begin{document}
\begin{frontmatter}
\title{\TITLE }
\author{Stefano Laporta}
\ead{stefano.laporta@pd.infn.it}  
\address{Dipartimento di Fisica e Astronomia, Universit\`a di Padova,
Istituto Nazionale Fisica Nucleare, Sezione di Padova,
Via Marzolo 8, I-35131 Padova, Italy}
\begin{abstract}
We have evaluated with 1100 digits of precision the
contribution of all the 891 mass-independent 4-loop Feynman diagrams contributing 
to the slope of the Dirac form factor in QED. 
The total  4-loop contribution is 
$$ m^2 F_1^{(4)'}(0) =
 0.886545673946443145836821730610315359390424032660064745{\ldots} 
\app^4 \ .
$$
We have fit a semi-analytical expression to the numerical value. The
expression contains harmonic polylogarithms of argument
$e^{\frac{i\pi}{3}}$, 
$e^{\frac{2i\pi}{3}}$,
$e^{\frac{i\pi}{2}}$, one-dimensional integrals of products of
complete elliptic integrals and six finite parts of master integrals, evaluated up to 4800 digits.
We show the correction
on the shift of the energy levels of the hydrogen atom due to the slope.
\end{abstract}
\end{frontmatter}
Recently in Ref.\citep{Laporta:2017okg} 
the 4-loop contribution to the electron $g$-$2$ in QED was
calculated numerically with very-high precision, and a semi-analytical fit was
obtained. In this companion paper we use the same techniques to
calculate the 4-loop QED contribution to the first derivative of
the Dirac form factor.  

In QED the amplitude for a vertex function can be written
\begin{equation}
(-ie)\bar{u}(p_1)\Gamma_{\mu}(p_1,p_2) u(p_2) = 
(-ie)\bar{u}(p_1)\left(
\gamma_{\mu}F_1(t) +  \frac{\sigma_{\mu\nu}}{2m} q_{\nu} F_2(t)
\right)u(p_2) \ ,
\end{equation}
where $m$ is the electron mass, $p_1$, $p_2$ and $q$ are the momenta of the electrons and the
photon, satisfying 
\begin{equation}
p_1^2= p_2^2=-m^2, \quad q=p_1-p_2, \quad t=-q^2\ .
\end{equation}
$F_1(t)$ and $F_2(t)$ are the Dirac and Pauli form factors.
At $t=0$, charge conservation implies that  
\begin{equation}
 F_1(0)=1 \ ,
\end{equation}
whereas  the value  of the Pauli form factor is the $g$-$2$ 
\begin{equation}
 F_2(0)=\frac{g-2}{2} \ .
\end{equation}
The quantity $\left. \frac{d}{dt} F_1(t)\right|_{t=0}=F_1'(0)$ is the slope of the
Dirac form factor.
The slope can be expanded perturbatively in powers of $\app$  
\begin{equation}
\label{eqserie}
m^2 F_1'(0)=\aql{1} \app +\aql{2} \app^2 +\aql{3} \app^3 +\aql{4} \app^4 +{\ldots} \ .
\end{equation}
The coefficient $A_1$ in \eqref{eqserie} is I.R. divergent: 
\begin{equation}
\label{a1ana}
\aql{1}= -\dfrac{1}{8}-\dfrac{1}{6\e}\ ;
\end{equation}
the divergence is due to the on-mass-shell
condition of the external electron\footnote{ 
In the calculation of the shift to energy levels due to the slope at one loop, 
the off-mass-shell condition has to be taken into account, and
this removes the divergence and gives rise to the Bethe
logarithms\citep{Erickson:1965a,Erickson:1965b}};
from two loops onward, the coefficients are finite.
The two-loop
and three-loop
coefficients are known in analytical
form~\citep{Barbieri:1972as,Barbieri:1972hn,Melnikov:1999xp}
\begin{align}
\label{a2ana}
\aql{2}=&
-\frac{4819}{5184} - \frac{49}{72}\Z2
- \frac{3}{4}\Z3
+ 3 \Z2 \ln2
=
0.469\;941\;487\;459\;992{\ldots}\ , 
\end{align}
\begin{align}
\label{a3ana}
\aql{3} =& 
   - \frac{77513}{186624}
   - \frac{454979}{6480}\Z2
   - \frac{2929}{288}\Z3
   + \frac{41671}{360}\Z2 \ln2
   + \frac{3899}{288}\Z4 
\cr &
   - \frac{103}{180}\Z2 \ln^2 2
   - \frac{217}{9} \left( \A4 + \frac{1}{24}\ln^4 2 \right) 
   + \frac{25}{8}\Z5
   - \frac{17}{4}\Z3\Z2
\cr &
=
0.171\;720\;018\;909\;775{\ldots} \ ,
\end{align}
where  $\zeta(n)=\sum_{i=1}^\oo i^{-n}$,
 $a_n=\sum_{i=1}^\oo 2^{-i}\;i^{-n}$. 

In this paper we present the result of the 
calculation of $\aql{4}$ with a precision of 1100 digits.
\begin{table}
\scriptsize
\begin{center}
\begin{tt}
0.8865456739464431458368217306103153593904240326600647453680559093208403164656289274548364863241773368\\
693512758747218307996875923974888466826147611753011917584831446774752672980326917402719214651539325519\\
844793100495019624531372119372946716080063429980958425369584945060683836659851413873218942100123948827\\
595153823786537220388349644856007568985761687756410271977960391029027661512235640610539922790515027760\\
822459236950433275703613350935251764763992516822679359645249285456658218441028674547644077579921118603\\
788315350119800677785150747802126742479040522224733029502183107429019902991627682916022890589911642646\\
344987898763072708284836435874347800245541537243400896951471683115538642559188352093478066512674887503\\
345902599182245563613125124119880615415537621337112284846277684867421928289686568115480303537276007873\\
036210930592647529598922340178357328289717496239918335278488413242436969926422136403200684400061242352\\
981583396633256675315824174144821761659738127669216197667509505074064930956136195898802456451163545675\\
716230944173884811565020098334847940590188785421700667378220853053541953188378610075518116338519220{\ldots}
\end{tt}
\end{center}
\caption{\CAPTIONTABLEA}
\label{tab:a1100}
\end{table}
The first digits of the result are 
\begin{equation}\label{ae4loop}
\aql{4}=
0.886545673946443145836821730610315359390424032660064745368055909{\ldots}\ . 
\end{equation}
The full-precision result is shown in table \ref{tab:a1100}.
\Figuraboh
We note that $\aql{2}$, $\aql{3}$ and $\aql{4}$ are all positive,
in contrast with the alternating signs observed in the
$g$-$2$ up to 5 loops.

Let us now consider the shift to the hydrogen energy levels  due to $\aql{4}$. 
We express the energy shift in terms of the frequency shift $\Delta f=\Delta E/h$.
For the level $n$S the frequency shift is~\citep{Erickson:1965a,Erickson:1965b}

\begin{equation}\label{an}
\Delta f_{\text{slope}}
(n\text{S, 4-loop})=
\frac{4
(Z\alpha)^4 m c^2}{h \;n^3}\left(\frac{m_r}{m}\right)^3 \left[ \app^4
A_4 \right] \ ,
\end{equation}
where $m_r$ is the reduced mass $m_r=m M/(m+M)$ and $M$ is the proton
mass.
Inserting the values of $m$, $M$, $c$, $h$ and $Z=1$, the
correction due to $\aql{4}$ is 
\begin{equation}\label{corrq}
\Delta f_{\text{slope}} (n\text{S, 4-loop}) =
%
\frac{
36.11 }{n^3}     \;\text{Hz} \ ,
\end{equation}
and is comparable with the experimental error of the extremely
precise measurement
of $1\text{S}-2\text{S}$ transition\citep{Matveev:2013orb}
\begin{equation}
f(1\text{S}-2\text{S})= 2466\; 061\; 413\; 187\; 018\pm 11 \;\text{Hz} \ .
\end{equation}
\eqref{corrq} is the first calculated 4-loop correction to energy
levels, of the kind  $\app^4(Z\alpha)^4$;
we note that there are some two-loop and three-loop radiative corrections
which still have theoretical errors larger than \eqref{corrq},
of the order of ~$10 \app^2(Z\alpha)^6$ and ~$10 \app^3(Z\alpha)^5$, respectively 
(see~\citep{Yerokhin:2018gna,Karshenboim:2019iuq}).

Now we consider the shift due to all the QED 4-loop contributions: 
$\aql{4}$ from $F_1'(0)$,
$\cql{4}$ from $g$-$2$ (see Eq.(2) of Ref.\citep{Laporta:2017okg}) and 
$\Pi^{(4)}_{le,1}$ from vacuum polarization (see Eq.(4) of
Ref.\citep{Baikov:2013ula}).
Writing  
\begin{equation}
\Delta f_{\text{4-loop QED}}(n\text{S})=
\frac{
(Z\alpha)^4 m c^2}{h \;n^3}\left(\frac{m_r}{m}\right)^3 \left[ \app^4
D_{40} \right]\ ,
\end{equation}
then 
\begin{align}\label{D40}
D_{40}=&4\aql{4}+\cql{4} - \Pi^{(4)}_{le,1}
= 
3.546182 -1.912245 -1.583612
\cr&
= 
0.05032465082590245550858429619942750274917{\ldots} \ . 
\end{align}
Note the deep numerical cancellation.
Therefore 
\begin{equation}
\Delta f_{\text{4-loop QED}} (n\text{S}) =
\frac{0.513}{n^3}     \;\text{Hz} \ .
\end{equation}

\Figuragau
There are 891 vertex diagrams contributing to $\aql{4}$.  They can be
obtained by inserting an external photon in each possible electron
line of the 104 4-loop self-mass diagrams shown in Fig.\ref{self104}.
Because of the Furry's theorem, the vertex diagrams with
closed electron loops with an odd number of vertices do not
contribute, and are not considered. The vertex diagrams can be
arranged in 25 gauge-invariant sets (Fig.\ref{figuragau}).
The sets are classified according to the number of photon corrections
on the same side of the main electron line and the insertions of
electron loops (see Ref.\citep{Cvitanovic:1977dp}). The numerical
contributions of each set, truncated to 40 digits, are listed in the
table \ref{tableset}.
Adding the contributions of diagrams with and without
closed electron loops one finds 
\begin{align} 
 \aql{4}(&\textrm{no closed electron loops})    =
 \cr
 &\qquad\qquad\qquad\qquad\phantom{+} 0.3514798015766637774090446716934794695266 \ ,\\
 \aql{4}(&\textrm{closed electron loops only})  =
 \cr
 &\qquad\qquad\qquad\qquad\phantom{+} 0.5350658723697793684277770589168358898637 \ .
\end{align}   

By building systems of 
integration-by-parts identities\citep{Chetyrkin:1981qh,Tkachov:1981wb}
and solving them\citep{Laporta:2001dd}, 
the contributions of all the diagrams to $\aql{4}$ are  expressed as
linear combinations of  334
master integrals, the same ones as appeared in the calculation of 4-loop $g$-$2$
\citep{Laporta:2017okg}. 

In Ref.\citep{Laporta:2017okg} these master integrals were calculated
numerically with precision ranging from 1100 to 9600 digits; 
analytical expressions were fit to all these master integrals 
(single or in particular combinations) 
by using the PSLQ algorithm\citep{PSLQ,Bailey:1999nv}.

For the scope of this work these results suffice, 
with the exception of a new combination of elliptic master integrals,
which has been successfully fit
by using the same basis used for the other master integrals. 

Therefore, 
the analytical expression of $\aql{4}$ contains 
the same transcendentals appeared in the $g$-$2$ result:
values of harmonic polylogarithms\citep{Remiddi:1999ew,Gehrmann:2001pz}
with argument $1$, $\frac{1}{2}$, $e^{\frac{i\pi}{3}}$ ,
$e^{\frac{2i\pi}{3}}$,  $e^{\frac{i\pi}{2}}$~\citep{Ablinger:2011te,Laporta:2018eos},
a family of one-dimensional integrals of products of
elliptic integrals, and the finite terms of the $\e-$expansions of six master integrals belonging to 
the topologies 81 and 83 of Fig.\ref{self104}. 
\begin{table}
\footnotesize
\begin{center}
\begin{tabular}{rrrr}
\hline
   1  & \phantom{+} 0.1350531726346435372674724541103838371038   \\
   2  & \phantom{+} 0.3802929165240844585552528298843579658371   \\
   3  &           - 0.0789488893676831608109628366941799823079   \\
   4  & \phantom{+} 0.3662786736588470044584250527325325702299   \\
   5  &           - 1.0979832148317652705103820073196531832520    \\
   6  & \phantom{+} 0.6467871429585372084492391789800382619165   \\
   7  & \phantom{+} 0.0895891170440342216099366534902414320652  \\
   8  &           - 0.3322086225106643608126657791889571079890   \\
   9  & \phantom{+} 0.0763376479373933425961220467893817339605  \\
   10 & \phantom{+} 0.2118669010888818123786340161652003594809   \\
   11 &           - 0.0541837571893361764657206136746826299854  \\
   12 & \phantom{+} 0.0108761535582321058694530867351119912448  \\
   13 &           - 0.0142646608196830116628021692409901716905  \\
   14 &           - 0.0058117416010420357833143542203438251011 \\
   15 &           - 0.2439068506475319592123409557076293747890   \\
   16 & \phantom{+} 0.2062012570841125786262218639260170000956   \\
   17 & \phantom{+} 0.0085366428673036656037790352019835488011 \\
   18 & \phantom{+} 0.0533927095302949341276880145918233326838  \\
   19 & \phantom{+} 0.0236058911191014021135877461122766184082  \\
   20 & \phantom{+} 0.0740163162205724051338179043210727390276  \\
   21 &           - 0.0537711607064956999082765338567906834199  \\
   22 & \phantom{+} 0.1819474273966664016975772159395176159307   \\
   23 & \phantom{+} 0.2359289294543601921365690660148707901595   \\
   24 &           - 0.0021225895319909487365222280699442649666 \\
   25 & \phantom{+} 0.0690362620755704991160330435886767859471  \\
\hline
\end{tabular}
\end{center}
\caption{\CAPTIONTABLEB}
\label{tableset}
\end{table}
The result of the analytical fit is written as follows:
\begin{align}\label{Tall}
\aql{4}=&T_0+T_2+T_3+T_4+T_5+T_6+T_7
 +\sqrt{3} \left( V_{4a}+V_{6a}\right)  +V_{6b}+V_{7b}
  \cr&
  +W_{4a}   +W_{6b} +W_{7b}
 +\sqrt{3} \left(E_{4a}+E_{5a}+E_{6a}+E_{7a}\right) +E_{6b}+E_{7b} +U\ .
\end{align}
The terms have been arranged in blocks with equal transcendental
weight. The index number is the weight.
The terms containing the ``usual'' transcendental constants are:
\begin{equation}\label{T023}
  T_0+T_2+T_3= 
          - \frac{92473962293}{19752284160}
          - \frac{6619898477}{21772800} \Z2
          - \frac{12334741}{132300} \Z3
          + \frac{97832509}{90720} \Z2 \ln 2 \ ,
\end{equation}	  
\begin{equation}\label{T4}
T_4=
          - \frac{241619904061}{391910400} \Z4
          + \frac{4572662443}{12247200} \Z2 \ln^2 2 
          - \frac{1449791143}{3061800} \comba{4}
          \ ,
\end{equation}	  
\begin{equation}\label{T5}
T_5=
            \frac{90355973}{134400} \Z5
          + \frac{1173056009}{9072000} \Z3 \Z2
          - \frac{8548241}{30240} \Z4 \ln 2 
          - \frac{68168}{135} \comba{5}\ ,
\end{equation}	  
\begin{align}\label{T6}
T_6=&
          - \frac{244603373713}{52254720} \Z6
          - \frac{8082848863}{24192000} \ZTD
          + \frac{159693503}{72000} \Z3 \Z2 \ln 2
\cr&
          - \frac{328317209}{302400} \Z4 \ln^2 2
          + \frac{402152509}{189000} \comba{4}\Z2 
          - \frac{18215}{27} \comba{{61}}
          + \frac{26062}{27} \comba{{62}}
	  \ ,
\end{align}	  
\begin{align}\label{T7}
T_7=&
          - \frac{7224951103}{1741824} \Z7
          - \frac{1267114025}{387072} \Z4 \Z3
          - \frac{2749470791}{387072} \Z5 \Z2
\cr&
          + \frac{971827}{128} \Z6 \ln 2
          - \frac{6242389}{6048} \Z3 \Z2 \ln^2 2
          - \frac{427145}{504}  \comba{4} \Z3 
\cr&
          + \frac{1420289}{180} \comba{5} \Z2  
          + \frac{256321}{756} \comba{{71}}
          - \frac{116987}{63} \comba{{72}}
          + \frac{104041}{20} \comba{{73}}
	  \ ,
\end{align}	  
where
\begin{align}\label{taa}
\comba{4}=&\A4+\frac{1}{24}\ln^4 2 \ ,\qquad
\comba{5}=\A5  + \frac{1}{12}\Z2\ln^3 2 -\frac{1}{120}\ln^5 2 \ ,
  \\
\comba{{61}}=&\B6- \A5 \ln 2 + \Z5 \ln 2 + \frac{1}{6}\Z3 \ln^3 2 -\frac{1}{12}\Z2\ln^4 2 + \frac{1}{144}\ln^6 2
\ ,
\\
\comba{{62}}=&\A6 -\frac{1}{48}\Z2 \ln^4 2 +\frac{1}{720}\ln^6 2
\ ,
\end{align}
\begin{align}
\comba{{71}}=& \D7 - 2\B6 \ln 2 + 4\A6\ln 2 + 2 \A5 \ln^2 2 -
\frac{49}{32}\ZTD \ln 2 -\frac{95}{32} \Z5 \ln^2 2
       +\frac{1}{8} \Z4  \ln^3 2
       \cr&
       -\frac{1}{3} \Z3  \ln^4 2
       +\frac{1}{12} \Z2  \ln^5 2 -\frac{1}{120} \ln^7 2 
	  \ ,
\end{align}
\begin{align}
\comba{{72}}=&\B7
 - 3 \A7 - \A6 \ln 2 -\frac{1}{2} \Z5  \ln^2 2 +\frac{1}{48} \Z4  \ln^3 2 -
 \frac{1}{24} \Z3 \ln^4 2 
\cr&
 +\frac{ 1}{120} \Z2 \ln^5 2
 -\frac{1}{1680} \ln^7 2 \ ,
\end{align}
\begin{align}
\comba{{73}}=&\left(
\A4-\frac{1}{4}\Z2 \ln^2 2  +\frac{ 7}{16} \Z3 \ln 2 +\frac{ 1}{24}
\ln^4 2
\right) \Z2 \ln 2 
\ .
\end{align}
The terms containing harmonic polylogarithms of $e^{\frac{i\pi}{3}}$, $e^{\frac{2i\pi}{3}}$:
\begin{equation}\label{V4a}
 V_{4a}=
          - \frac{14186171}{194400} \Cl4
          - \frac{103023803}{583200} \Z2 \Cl2\ ,
\end{equation}	  
\begin{equation}\label{V6a}
 V_{6a}=
            \frac{916598}{76545} \combp{{61}}
          + \frac{844343}{28350} \combp{{62}}
          + \frac{178619489}{3980340} \combp{{63}}
          - \frac{263673944}{295245} \combp{{64}} \ ,
\end{equation}	  
\begin{equation}\label{V6b}
 V_{6b}=
            \frac{212671}{2400} \combp{{65}}
          - \frac{1031987}{14400}  \Z2 \Cldd\ ,
\end{equation}	  
\begin{equation}\label{V7b}
 V_{7b}=
          - \frac{507}{4} \combp{{71}}
          - \frac{295}{4} \combp{{72}} \ ,
\end{equation}	  
where
\begin{align}
\combp{{61}}=& \iha{0,0,0,1,-1,-1} 
       + \ihb{0,0,0,1,-1,1}
       + \ihb{0,0,0,1,1,-1}
\cr&
       + \frac{27}{26}\ihb{0,0,1,0,1,1}
       + \frac{207}{104}\ihb{0,0,0,1,1,1}
       + \frac{10}{3}\A4 \Cl2
         \cr&
       + \frac{7}{4} \Z3 \iha{0,1,-1}
       + \frac{21}{8} \Z3 \ihb{0,1,1}
       - \frac{5}{72} \Z3 \Z2 \pi
       \cr&
       - \frac{5}{6} \Cl2 \Z2\ln^2 2
       + \frac{5}{36} \Cl2 \ln^4 2
       - \frac{27413}{67392} \Z5 \pi
       \cr&
       + \frac{4975}{11583} \Z4 \Cl2
	  \ ,
\end{align}
\begin{align}
\combp{{62}}& = \Z2 \biggl(
         \iha{0,1,1,-1} 
       + \frac{3}{2} \ihb{0,1,1,-1}
       - \frac{1}{6} \Z3 \pi
       + \frac{1}{108} \Z2 \pi \ln 2
       \cr&
       - \frac{5}{2} \iha{0,1,-1} \ln 2
       - \frac{15}{4} \ihb{0,1,1} \ln 2
       + \frac{25}{12} \Cl2 \ln^2 2
       \cr&
       - \frac{661}{1188} \Cl2 \Z2
       \biggr)
	  \ ,
\end{align}
\begin{align}
\combp{{63}}=&\Cl6 - \frac{3}{4} \Z4 \Cl2 \ , \qquad      
\combp{{64}} = \Cl4 \Z2 - \frac{91}{66} \Z4 \Cl2  \ ,
\\
\combp{{65}}=&\rha{0,0,0,1,0,1}+\Cl2 \Cl4\ ,
\end{align}
\begin{align}
\combp{{71}}=&\rha{0,0,0,1,0,1,-1} 
       + 4 \rha{0,0,0,0,1,1,-1}
       - \frac{27}{8} \rhb{0,0,1,0,0,1,1}
\cr&
       - \frac{135}{16} \rhb{0,0,0,1,0,1,1}
       - \frac{27}{2} \rhb{0,0,0,0,1,1,1}
\cr&
       + \iha{0,1,-1} \Cl4
       + \frac{3}{2} \ihb{0,1,1} \Cl4
       + \frac{145}{132} \Cl6 \pi 
       \ ,
\end{align}
\begin{align}
\combp{{72}}=& \Z2 \biggl(
         \rha{0,1,0,1,-1}
       + 2 \rha{0,0,1,1,-1}
       + \frac{9}{4} \rhb{0,1,0,1,1}
\cr&
       + \frac{9}{2} \rhb{0,0,1,1,1}
       + \iha{0,1,-1} \Cl2
\cr&
       + \frac{3}{2} \ihb{0,1,1} \Cl2
       \biggr)
       \ .
\end{align}
The terms containing harmonic polylogarithms of $e^{\frac{i\pi}{2}}$:
\begin{align}
\label{W4a}
 W_{4a}=&
          - \frac{1117}{36}  \Z2 \cat2  \ ,
   \\\cr
\label{W6b}
 W_{6b}=&
           \frac{38424}{125}  \Z2 \catdd \ ,
   \\\cr
\label{W7b}
 W_{7b}=&
          - 472 \combp{{73}}
	  \ ,
\end{align}	  
where
\begin{align}
\combp{{73}}=&\Z2 \biggl(
      \rhe{0,1,0,1,1}
       + \cat2 \ihe{0,1,1}
       - \frac{1}{2} \cat4 \pi
\cr&
       + \frac{1}{4} \catdd  \ln 2
       \biggr)
       \ .
\end{align}
A term $\Z2\cat2$ appears in  \eqref{W4a};
it did not appear  in the 4-loop $g$-$2$ result\citep{Laporta:2017okg}
because of cancellations in the final sum of all 4-loop diagrams.
The terms containing elliptic constants:
\begin{equation}\label{E4a}
 E_{4a}=
        \pi\left(
            \frac{5581729229}{362880000} \btell
          + \frac{1233637481}{1399680000} \ctell
	  \right) \ ,
\qquad
 E_{5a}=
          - \frac{11495611}{3265920}  \pi \intaxonetwos{0,0,1}  \ ,
\end{equation}	  
\begin{equation}\label{E6ab}
 E_{6a}=
          - \frac{365478661}{24494400} \combel{{61}}
          + \frac{119022487}{5443200} \combel{{62}}\ ,
\qquad
 E_{6b}=
          - \frac{751}{729} \Z2 \intaxoneones{0,0,1}\ ,
\end{equation}	  
\begin{equation}\label{E7ab}
 E_{7a}=
          - \frac{98285}{248832} \combel{{71}}
          - \frac{157753}{497664} \combel{{72}}\ ,
\qquad
 E_{7b}=
            \frac{157753}{41472} \combel{{73}}
          - \frac{99731}{1944} \combel{{74}} \ ,
\end{equation}	  
where
\begin{align}
\combel{{61}}=& \pi \left(
   \intaxonetwos{0,2,0}
 - \frac{9}{4} \ln 2 \intaxonetwos{0,0,1}
 \right)  
       \ ,
\\
\combel{{62}}=& \pi \left(
         \intaxonetwos{0,1,1}
       - \frac{3}{8} \intaxonetwos{0,0,2}
       - \frac{3}{2} \ln 2 \intaxonetwos{0,0,1}
       \right)
       \ ,
\end{align}
\begin{align}
\combel{{71}}=&\pi \biggl(
          \intaxonetwos{2,1,0}
       + \frac{7}{3} \intaxonetwos{1,2,0}
       - 2 \intaxonetwos{1,1,1}
       + \frac{40}{27} \intaxonetwos{0,3,0}
       - \frac{7}{3} \intaxonetwos{0,2,1}
\cr&      
       + \intaxonetwos{0,1,2}
       - 30 \ln 2 \intaxonetwos{0,2,0}
       + 45 \ln 2 \intaxonetwos{0,1,1}
       - \frac{135}{8} \ln 2 \intaxonetwos{0,0,2}
       \biggr) 
       \ ,
\end{align}
\begin{align}
\combel{{72}}=&\pi \biggl(
          \intaxonetwos{2,0,1}
       + \frac{14}{3} \intaxonetwos{1,2,0}
       - 2 \intaxonetwos{1,1,1}
       - 2 \intaxonetwos{1,0,2}
       - \frac{370}{27} \intaxonetwos{0,3,0}
\cr&      
       + \frac{85}{3} \intaxonetwos{0,2,1}
       - 22 \intaxonetwos{0,1,2}
       + 7 \intaxonetwos{0,0,3}
       + 11 \Z2 \intaxonetwos{0,0,1}
\cr&      
       - 20 \ln 2 \intaxonetwos{0,2,0}
       + 30 \ln 2 \intaxonetwos{0,1,1}
       - \frac{45}{4} \ln 2 \intaxonetwos{0,0,2}
       \biggr)
       \ ,
\end{align}
\begin{align}
\combel{{73}}=&\Z2 \left(
         \intaxoneones{1,0,1}
       - \intaxoneones{0,1,1}
       + \frac{1}{4} \intaxoneones{0,0,2}
       \right)
       \ ,
\\      
\combel{{74}}=&\Z2 \left(\label{ezz}
         \intaxoneones{0,2,0}
       - \frac{3}{2} \intaxoneones{0,1,1}
       + \frac{9}{16} \intaxoneones{0,0,2}
       \right)  
       \ .
\end{align}

The term containing the  $\e^0$ coefficients of the $\e-$expansion of
six master integrals (see $f$, $f'$, $f''$, $g$, $g'$, $g''$ of Fig.\ref{unkmasd}):
\begin{equation}
   U =\label{UU}
        \frac{174623}{288000} C_{81a}
       +\frac{29479}{7200}    C_{81b}
       -\frac{43}{6}          C_{81c}
       +\frac{10871}{14400}   C_{83a}
       -\frac{157}{1620}      C_{83b}
       -\frac{95}{24}         C_{83c}\ .
\end{equation}	  
\Figuraunkmasd

In the above expressions  
 $b_6=H_{0,0,0,0,1,1}\left(\frac{1}{2}\right)$,  
 $b_7=H_{0,0,0,0,0,1,1}\left(\frac{1}{2}\right)$, 
 \\
 $d_7=H_{0,0,0,0,1,-1,-1}(1)$, 
 $\cl{n}{\theta}=\Im \Li_n (e^{i\theta})$.
$H_{i_1,i_2,{\ldots} }(x)$ are the harmonic polylogarithms.
The integrals $f_j$ are defined as follows:
\begin{align}\label{fdef}
f_m(i,j,k)=&\int_1^9 ds\; D_1(s) \Re\left(\sqrt{3^{m-1}} D_m(s)\right)  \left(s-\frac{9}{5}\right)
\ln^i\left(9-s\right)
\ln^j\left(s-1\right)
\ln^k\left(s\right)
\ ,
\\\cr
D_m(s)=&\frac{2}{\sqrt{(\rs+3)(\rs-1)^3}}K\left(m-1-(2m-3)\frac{(\rs-3)(\rs+1)^3}{(\rs+3)(\rs-1)^3}\right)\
;
\end{align}
$K(x)$ is the complete elliptic integral of the first kind.
The constants 
$B_3$ and $C_3$ have
the following hypergeometric representations~\citep{Laporta:2008sx,Zhou:2019aa}:
\begin{align}\label{cdef}
B_3=\int_0^1 dx &\dfrac{K_c^2(x)}{\sqrt{1-x}} =
\dfrac{\pi}{27}\sqrt{3}\left(
{}_4\WF_3\left(\begin{smallmatrix}
{{\frac{1}{6}\;\frac{1}{3}\;\frac{1}{3}\;\frac{1}{2}}}\\
{{\frac{5}{6}\;\frac{5}{6}\;\frac{2}{3}}}\end{smallmatrix}; 1\right)
-
{}_4\WF_3\left(\begin{smallmatrix}
{{\frac{5}{6}\;\frac{2}{3}\;\frac{2}{3}\;\frac{1}{2}}}\\
{{\frac{7}{6}\;\frac{7}{6}\;\frac{4}{3}}}\end{smallmatrix}; 1\right)
\right) \ , \\\cr
C_3=\int_0^1 dx &\dfrac{E_c^2(x)}{\sqrt{1-x}} =
\dfrac{\pi}{27}\sqrt{3}\left(
{}_4\WF_3\left(\begin{smallmatrix}
{{\frac{1}{6}\;\frac{1}{3}\;\frac{4}{3}\;-\frac{1}{2}}}\\
{{-\frac{1}{6}\;\frac{5}{6}\;\frac{5}{3}}}\end{smallmatrix}; 1\right)
-
{}_4\WF_3\left(\begin{smallmatrix}
{{-\frac{7}{6}\;-\frac{1}{3}\;\frac{2}{3}\;-\frac{1}{2}}}\\
{{-\frac{5}{6}\;\frac{1}{6}\;\frac{1}{3}}}\end{smallmatrix}; 1\right)
\right) \ , 
\end{align}
\begin{align}
{}_4\WF_3\left(\begin{smallmatrix}
{{a_1\;a_2\;a_3\;a_4}}\\
{{b_1\;b_2\;b_3}}\end{smallmatrix}; x\right)
&
=\dfrac{\Gamma{(a_1)}\Gamma{(a_2)}\Gamma{(a_3)}\Gamma{(a_4)}}{\Gamma{(b_1)}\Gamma{(b_2)}\Gamma{(b_3)}}
{}_4F_3\left(\begin{smallmatrix}
{{a_1\;a_2\;a_3\;a_4}}\\
{{b_1\;b_2\;b_3}}\end{smallmatrix}; x\right)\ ,
\\\cr
K_c(x)=&\frac{2\pi}{\sqrt{27}} \Fcub{x} \ , \qquad E_c(x)=\frac{2\pi}{\sqrt{27}} \Ecub{x} \ . 
\end{align}
\begin{table}
\footnotesize
\begin{center}
\begin{tabular}{rrrr}
\hline
$T_0$    &    - 4.681684484889468094812989972699947224\\
$T_2$    &  - 500.133034055977659141446135278499933321\\
$T_3$    &   1117.500891445130805660768567602555268503\\
$T_4$    &  - 621.782936431818861978090058789852717683\\
$T_5$    &    461.630448606722732032079258609728703085\\
$T_6$    &  - 722.889129056625650906898067019971503340\\
$T_7$    & - 1920.880025680053685984763498010088725117\\
$V_{4a}$ &  - 361.756789173538133855596918541497826293\\
$V_{6a}$ &   - 12.795973342316846821724756996806906345\\
$V_{6b}$ &   - 43.243682435714549680745107097288577324\\
$V_{7b}$ &  - 357.420812721946242890859711585624260199\\
$W_{4a}$ &   - 46.749646168999285541674809967204826369\\
$W_{6b}$ &    424.228046686093592380691247689052425405\\
$W_{7b}$ &   1161.850798649722146670796323365721460341\\
$E_{4a}$ &    363.984514808233148461875123331608293748\\
$E_{5a}$ &  - 340.007389863188265877938835467376960664\\
$E_{6a}$ &    282.390876991327380717000994357591349928\\
$E_{6b}$ &   - 28.367551495832530043307768783586340006\\
$E_{7a}$ &  - 460.255472174720354400490351968210139167\\
$E_{7b}$ &   1956.590087984274945528724322574247574302\\
$U$      &     40.520106727663067648614920217821543664\\
$C_{81a}$ &   116.694585791186600526332510987652818034\\
$C_{81b}$ &   - 8.748320323814631572671010051472284815\\
$C_{81c}$ &   - 0.236085277120339887503638687666535683\\
$C_{83a}$ &     2.771191986145520146810618363218497216\\
$C_{83b}$ &   - 0.807847353263827557176395243854200179\\
$C_{83c}$ &   - 0.434702618543809180642530601495074086\\
\hline
\end{tabular}
\end{center}
\caption{\CAPTIONTABLEC}
\label{tableter}
\end{table}
The numerical values of the constants appearing in \eqref{Tall}
are listed in Table \ref{tableter}.
The right-hand sides of 
\eqrefb{T4}{T7},
\eqrefb{V6a}{V7b},
\eqref{W7b} and 
\eqrefb{E6ab}{E7ab}
have been written by using
some suitable combinations of constants, 
$\comba{i}$, $\combp{i}$ and $\combel{i}$, found by comparing the
fits of several contributions of diagrams to $\aql{4}$ and $F_2(0)$.
In this way, we obtain a decomposition of $\aql{4}$ as linear
combinations of the elements of a basis of only $57$ objects (the
terms in the right-hand sides of  
\eqrefb{T023}{T7},
\eqrefb{V4a}{V7b},
\eqrefb{W4a}{W7b},
\eqrefb{E4a}{E7ab} and
\eqref{UU}).
We have found that \emph{each one} of the 891
contributions of the 4-loop vertex diagrams
to $F_1'(0)$ and to $F_2(0)$ can be written as linear combination of
the elements of this basis. 


We briefly describe the method used to obtain $\aql{4}$.
It is the same used in Ref.\citep{Laporta:2017okg}.
The 104 self-mass diagrams are generated with a $C$ program.
The contribution to $\aql{4}$ from the amplitude $M_{\mu}(p+q/2,p-q/2,q)$ of a 
vertex diagram is extracted by using projectors~\citep{Barbieri:1978it,Laporta:1900zz}
\begin{align}\label{proj}
F_1'(0)={\rm Tr} \biggl(
 &P^{(2)}_{\mu\nu}(p) \left.\frac{\partial M_{\mu}(p+q/2,p-q/2,q)}{\partial
 q_{\nu}}\right|_{q=0}
\cr
+&P^{(3)}_{\mu\nu\rho}(p) \left. \frac{\partial^2 M_{\mu}(p+q/2,p-q/2,q)}{\partial
q_{\nu}\partial q_{\rho}}\right|_{q=0}
\biggr) \ , 
\end{align}
analogously to the corresponding formula for $g$-$2$
\begin{equation}\label{projf2}
F_2(0)={\rm Tr} \biggl(
 P^{(0)}_{\mu}(p) M_{\mu}(p,p,0)
+P^{(1)}_{\mu\nu}(p) \left. \frac{\partial M_{\mu}(p+q/2,p-q/2,q)}{\partial
q_{\nu}}\right|_{q=0}
\biggr) \ ; 
\end{equation}
we use a {\tt FORM}\citep{FORM,Kuipers:2012rf} program to
perform this operation.
For each self-mass diagram a large system of 
integration-by-parts identities\citep{Chetyrkin:1981qh,Tkachov:1981wb}
is generated and solved by using the program{ \SYS}\citep{Laporta:2001dd}.
Using this system of identities the contribution of each diagram is
reduced to master integrals, which are the same  of Ref.\citep{Laporta:2017okg}.

The contribution of a diagram to the slope must be independent of the
internal routing chosen for the external momentum of the photon $q$.
We compute the
contributions with two different routings, one minimizing and the
other maximizing the number of momenta containing $q$. We check that both
expressions are reduced to same combination of master integrals.

Let us compare the contributions to the slope and to $g$-$2$ of the
same diagrams.
Due to the second derivative appearing in \eqref{proj},  
the contribution to the slope contains Feynman integrals with 
sum of exponents increased by 2 in the numerators and increased by 1 in
the denominators.
The total number of Feynman integrals
of a contribution increases typically of a factor $\sim 10-20$. 

For the same reason, the  number of identities of the system 
necessary to
reduce the contributions to the slope increases of a factor $10-30$ (up to
~$5\times 10^8$), and the size increases of a factor 10 (up to ~$1.5$TB).

For example, let us consider the contributions 
from the vertex diagrams derived from the
self-mass diagram 22 of Fig.\ref{self104};
in the sector with all the 11 denominators 
the Feynman integrals have 
maximum  sum of the exponents of the scalar products equal to 7, and 
maximum  sum of the exponents of the denominators minus the number
of denominators equal to 3.
The integrals which have maximum sum of exponents are generated 
by the derivative with respect to the external photon momentum; 
we have verified that it is not necessary to generate integration-by-parts 
identities 
which contain Feynman integrals with total sum of exponents greater than
these maxima.

\setcounter{secnumdepth}{0} 
\section{Acknowledgments}
This work has been supported by the Supporting TAlent in ReSearch at
Padova University (UniPD STARS Grant 2017 ``Diagrammalgebra'').

I wish to thank Pierpaolo Mastrolia for the encouragement and the support.
I wish to thank Thomas Gehrmann for providing me the access to the computing facilities
of the Institute for Theoretical Physics of Zurich, where 
most of the calculations was performed; 
the remaining part was performed on the CloudVeneto infrastructure 
of the Department of Physics and Astronomy and INFN of Padua.

I wish to acknowledge the organizers of RADCOR 2019, where the results
of this work were presented, for the invitation and the support.


\end{document}